\documentclass{article}
\usepackage{graphicx}
\usepackage{amsmath}

\begin{document}

\noindent \textbf{Comment on ``Torque or no torque? Simple charged particle
motion }

\noindent \textbf{observed} \textbf{in different inertial frames,'' by J. D.
Jackson [Am. }

\noindent \textbf{J. Phys. 72 (12), 1484-1487 (2004)]}\medskip \bigskip

In this paper it is shown that the real cause of the apparent electrodynamic
paradox discussed by Jackson [J. D. Jackson, Am. J. Phys. \textbf{72}, 1484
(2004)] is the use of three-dimensional (3D) quantities $\mathbf{E}$, $%
\mathbf{B}$, $\mathbf{F}$, $\mathbf{L}$, $\mathbf{N}$, .. . When \emph{4D
geometric quantities} are used then there is no paradox and the principle of
relativity is naturally satisfied. \medskip \bigskip

In a recent paper in this Journal Jackson$^{1}$ discussed the apparent
paradox of different mechanical equations for force and torque governing the
motion of a charged particle in different inertial frames. Two inertial
frames $S$ (the laboratory frame) and $S^{\prime }$ (the moving frame) are
considered (they are $K$ and $K^{\prime }$ respectively in Jackson's
notation). In $S^{\prime }$ a particle of charge $q$ and mass $m$
experiences only the radially directed electric force caused by a point
charge $Q$ fixed permanently at the origin. Consequently both $\mathbf{L}%
^{\prime }$ and the torque $\mathbf{N}^{\prime }$ are zero in $S^{\prime }$,
see Fig. 1(a) in Ref. 1. (The vectors in the three-dimensional (3D) space
will be designated in bold-face.) In $S$ the charge $Q$ is in uniform motion
and it produces \emph{both} an electric field $\mathbf{E}$ and a \emph{%
magnetic field} $\mathbf{B}$. The existence of $\mathbf{B}$ in $S$ is
responsible for the existence of the 3D magnetic force $\mathbf{F}=q\mathbf{V%
}\times \mathbf{B}$ and this force provides a 3D torque $\mathbf{N}$ ($%
\mathbf{N}=\mathbf{x}\times \mathbf{F}$) on the charged particle, see Fig.
1(b) in Ref. 1. Consequently a nonvanishing angular momentum of the charged
particle changes in time in $S$, $\mathbf{N}=d\mathbf{L}/dt$. Here we repeat
Jackson's words$^{1}$ about such result: ``How can there be a torque and so
a time rate of change of angular momentum in one inertial frame, but no
angular momentum and no torque in another? Is there a paradox? Some
experienced readers will see that there is no paradox - \emph{that is just
the way things are}, ...'' (my emphasis) Such reasoning is considered to be
correct by many physicists. However in the considered case \emph{the
principle of relativity is violated} and the ``explanation'' of the type ``%
\emph{that is just the way things are}'' does not remove the violation of
the principle of relativity.

In this Comment it will be shown that - \emph{it is not the way things are},
but that there is a simple solution of the above problem which is in a
complete accordance with the principle of relativity. The real cause of the
paradox is - \emph{the use of 3D quantities}, e.g., $\mathbf{E}$, $\mathbf{B}
$, $\mathbf{F}$, $\mathbf{L}$, $\mathbf{N}$, \emph{their transformations and
equations with them.} Instead of using 3D quantities we shall deal from the
outset with \emph{4D geometric quantities and equations with them}. In such
treatment the paradox does not appear and the principle of relativity is
naturally satisfied. The whole consideration is presented in much more
details in Ref. 2 in which the resolution of the paradox is exposed in four
different ways, whereas for this Comment we choose only one of them. There
(Ref. 2) we have also discussed the Trouton-Noble experiment in which
exactly the same paradox appears. It was shown$^{2}$ that the approach with
4D geometric quantities is in a complete agreement with experiment.

This investigation will be done in the geometric algebra formalism, which is
recently nicely presented in this Journal by Hestenes.$^{3}$ Physical
quantities will be represented by geometric 4D quantities, multivectors that
are defined without reference frames, i.e., as absolute quantities (AQs) or,
when some basis has been introduced, they are represented as 4D
coordinate-based geometric quantities (CBGQs) comprising both components and
a basis. For simplicity and for easier understanding only the standard basis
\{$\gamma _{\mu };0,1,2,3$\} of orthonormal 1-vectors, with timelike vector $%
\gamma _{0}$ in the forward light cone, will be used. For all mathematical
details regarding the spacetime algebra reader can consult Hestenes' paper.$%
^{3}$

Let us start the resolution of the paradox discussed by Jackson$^{1}$
writing all quantities as 4D AQs. The equations with them will be \emph{%
manifestly Lorentz invariant} equations. Thus the position 1-vector in the
4D spacetime is $x$. Then $x=x(\tau )$ determines the history of a particle
with proper time $\tau $ and proper velocity $u=dx/d\tau $. The Lorentz
force as a 4D AQ (1-vector) is $K_{L}=(q/c)F\cdot u$, where $u$ is the
velocity 1-vector of a charge $q$ (it is defined to be the tangent to its
world line).

The bivector field $F(x)$ (i.e., the electromagnetic field $F(x)$) for a
charge $Q$ with constant velocity $u_{Q}$ (1-vector) is
\begin{equation}
F(x)=kQ(x\wedge (u_{Q}/c))/\left| x\wedge (u_{Q}/c)\right| ^{3},  \label{cvf}
\end{equation}
where $k=1/4\pi \varepsilon _{0}$, see Ref. 2 and my references therein.
(For the charge $Q$ at rest, $u_{Q}/c=\gamma _{0}$.) All AQs in Eq. (\ref
{cvf}) can be written as CBGQs in some basis. We shall write them in the
standard basis $\{\gamma _{\mu }\}$. In the $\{\gamma _{\mu }\}$ basis $%
x=x^{\mu }\gamma _{\mu }$, $u_{Q}=u_{Q}^{\mu }\gamma _{\mu }$, $%
F=(1/2)F^{\alpha \beta }\gamma _{\alpha }\wedge \gamma _{\beta }$; the basis
components $F^{\alpha \beta }$ are determined as $F^{\alpha \beta }=\gamma
^{\beta }\cdot (\gamma ^{\alpha }\cdot F)=(\gamma ^{\beta }\wedge \gamma
^{\alpha })\cdot F$. In Hestenes' paper$^{3}$ the spacetime split is used
for the decomposition of $F$ into the electric and magnetic fields that are
represented by bivectors, see Eqs. (58)-(60) in Ref. 3. This means that
Hestenes' decomposition is \emph{an} \emph{observer dependent decomposition}%
; an observer independent quantity $F$ is decomposed into \emph{observer
dependent }bivectors of the electric and magnetic fields.

Instead of using the \emph{observer dependent decomposition} from Ref. 3 we
shall make an analogy with the tensor formalism$^{4}$ and represent the
electric and magnetic fields by 1-vectors $E$ and $B$ that are defined
without reference frames, i.e., as AQs. Thence they are \emph{independent of
the chosen reference frame and of the chosen system of coordinates in it}.
\begin{align}
F& =(1/c)E\wedge v+(IB)\cdot v,  \notag \\
E& =(1/c)F\cdot v,\quad B=-(1/c^{2})I(F\wedge v),  \label{itf}
\end{align}
where $I$ is the unit pseudoscalar. ($I$ is defined algebraically without
introducing any reference frame, as in Ref. 5, Sec. 1.2.) The velocity $v$
and all other quantities entering into the relations (\ref{itf}) are AQs.
That velocity $v$ characterizes some general observer. We can say, as in
tensor formalism,$^{4}$ that $v$ is the velocity (1-vector) of a family of
observers who measures $E$ and $B$ fields. Of course \emph{the relations for}
$E$ \emph{and }$B$, Eq. (\ref{itf}) \emph{hold for any observer;} they are
manifestly Lorentz \emph{invariant }equations. Note that $E\cdot v=B\cdot
v=0 $, which yields that only three components of $E$ and three components
of $B$ are independent quantities.

The 1-vectors $E$ and $B$ for a charge $Q$ moving with constant velocity $%
u_{Q}$ can be determined from (\ref{itf}) and the expression for the
bivector field $F$ (\ref{cvf}). They are
\begin{eqnarray}
E &=&(D/c^{2})[(u_{Q}\cdot v)x-(x\cdot v)u_{Q}]  \notag \\
B &=&(-D/c^{3})I(x\wedge u_{Q}\wedge v),  \label{ec}
\end{eqnarray}
where $D=kQ/\left| x\wedge (u_{Q}/c)\right| ^{3}$. When the world lines of
the observer and the charge $Q$ coincide, $u_{Q}=v$, then (\ref{ec}) yields
that $B=0$ and only an electric field (Coulomb field) remains.

The Lorentz force as a 4D AQ $K_{L}=(q/c)F\cdot u$ can be written in terms
of 4D AQs 1-vectors $E$ and $B$ as
\begin{equation}
K_{L}=(q/c)\left[ (1/c)E\wedge v+(IB)\cdot v\right] \cdot u.  \label{KEB}
\end{equation}
The equivalent expression in the tensor formalism, \emph{with tensors as AQs}%
, is recently given in this Journal, Ref. 4. Particularly from the
definition of the Lorentz force $K_{L}$ and the relation $E=(1/c)F\cdot v$
(from (\ref{itf})) it follows that the Lorentz force ascribed by an observer
comoving with a charge, $u=v$, is \emph{purely electric }$K_{L}=qE$. When $%
K_{L}$ is written as a CBGQ in $S$ and in the $\{\gamma _{\mu }\}$ basis it
is given as
\begin{equation}
K_{L}=(q/c^{2})[(v^{\nu }u_{\nu })E^{\mu }+\widetilde{\varepsilon }_{\ \nu
\rho }^{\mu }u^{\nu }cB^{\rho }-(E^{\nu }u_{\nu })v^{\mu }]\gamma _{\mu },
\label{lo}
\end{equation}
where $\widetilde{\varepsilon }_{\mu \nu \rho }\equiv \varepsilon _{\lambda
\mu \nu \rho }v^{\lambda }$ is the totally skew-symmetric Levi-Civita
pseudotensor induced on the hypersurface orthogonal to $v$.

Further the angular momentum $M$ (bivector), the torque $N$ (bivector) for
the Lorentz force $K_{L}$ and manifestly Lorentz invariant equation
connecting $M$ and $N$ are defined as
\begin{eqnarray}
M &=&x\wedge p,\ p=mu,  \notag \\
N &=&x\wedge K_{L};\quad N=dM/d\tau .  \label{MKN}
\end{eqnarray}
When $M$ and $N$ are written as CBGQs in the $\{\gamma _{\mu }\}$ basis they
become
\begin{eqnarray}
M &=&(1/2)M^{\mu \nu }\gamma _{\mu }\wedge \gamma _{\nu },\ M^{\mu \nu
}=m(x^{\mu }u^{\nu }-x^{\nu }u^{\mu }),  \notag \\
N &=&(1/2)N^{\mu \nu }\gamma _{\mu }\wedge \gamma _{\nu },\ N^{\mu \nu
}=x^{\mu }K_{L}^{\nu }-x^{\nu }K_{L}^{\mu }.  \label{mn}
\end{eqnarray}
We see that the components $M^{\mu \nu }$ from (\ref{mn}) are identical to
the covariant angular momentum four-tensor given by Eq. (A3) in Jackson's
paper.$^{1}$ However $M$ and $N$ from (\ref{MKN}) are 4D geometric
quantities, the 4D AQs, which are \emph{independent of the chosen reference
frame and of the chosen system of coordinates in it}, whereas the components
$M^{\mu \nu }$ and $N^{\mu \nu }$ that are used in the usual covariant
approach, e.g., Eq. (A3) in Ref. 1, are coordinate quantities, the numbers
obtained in the specific system of coordinates, i.e., in the $\{\gamma _{\mu
}\}$ basis. Notice that, in contrast to the usual covariant approach, $M$
and $N$ from (\ref{mn}) are also 4D geometric quantities, the 4D CBGQs,
which contain both components \emph{and a basis}, here bivector basis $%
\gamma _{\mu }\wedge \gamma _{\nu }$.

Let us now assume that the laboratory frame $S$ is the $\gamma _{0}$-system.
Thus in $S$ the observers who measure the fields are at rest, i.e., $%
v=v^{\mu }\gamma _{\mu }=c\gamma _{0}$, $v^{\mu }=(c,0,0,0)$. Then from (\ref
{itf}) it follows that in $S$ the temporal components of the 4D $E$ and $B$
are zero and only their spatial components remain. In the laboratory frame $S
$ both charges $Q$ and $q$ are moving and the components in the CBGQs $%
u_{Q}^{\mu }\gamma _{\mu }$ and $u^{\mu }\gamma _{\mu }$ are given as $%
u_{Q}^{\mu }=u^{\mu }=(\gamma c,\gamma \beta c,0,0)$. The fields $E$ and $B$
as AQs are given by (\ref{ec}) and when they are written as CBGQs in $S$
then the components $E^{\mu }$ become $E^{0}=E^{3}=0,$ $E^{1}=D\gamma
(x^{1}-\beta x^{0}),$ $E^{2}=D\gamma x^{2}$. Taking into account that in $%
S^{\prime }$ $t^{\prime }=0$, i.e., $x^{\prime 0}=\gamma (x^{0}-\beta
x^{1})=0$, the relation $x^{0}=\beta x^{1}$ is obtained. Inserting this last
relations into expressions for $E^{\mu }$ we find
\begin{equation}
E^{0}=E^{3}=0,\ E^{1}=Dx^{1}/\gamma ,\ E^{2}=D\gamma x^{2}.  \label{le}
\end{equation}
The charge\textbf{\ }$Q$ moves in the $S$ frame, which yields that the
magnetic field $B=B^{\mu }\gamma _{\mu }$ is now different from zero. The
components $B^{\mu }$ are
\begin{equation}
B^{0}=B^{1}=B^{2}=0,\ B^{3}=(1/c)D\gamma \beta x^{2}=\beta E^{2}/c.
\label{mg}
\end{equation}
The spatial components $E^{i}$ and $B^{i}$ from (\ref{le}) and (\ref{mg})
are the same as the usual expressions for the components of the 3D vectors $%
\mathbf{E}$ and $\mathbf{B}$. Inserting (\ref{le}) and (\ref{mg}) into (\ref
{lo}) we find the expression for the Lorentz force $K_{L}$ in the laboratory
frame $S.$ The components of $K_{L}$ in $S$ are
\begin{eqnarray}
K_{L}^{0} &=&q\gamma \beta E^{1},\ K_{L}^{1}=q\gamma E^{1},  \notag \\
K_{L}^{2} &=&q\gamma (E^{2}-\beta cB^{3})=qE^{2}/\gamma ,\ K_{L}^{3}=0.
\label{si}
\end{eqnarray}
We see that in $S$, when it is the $\gamma _{0}$-system in which the
observers who measure the fields are at rest, $v=c\gamma _{0}$, there is the
4D magnetic field (\ref{mg}) which enters into the expression for the total
4D Lorentz force $K_{L}$. Then using (\ref{le}), (\ref{mg}), (\ref{si}) and
the relation $x^{0}=\beta x^{1}$ one easily finds that all components $%
N^{\mu \nu }$ are zero
\begin{eqnarray}
x^{3} &=&0,\ K_{L}^{3}=0\Rightarrow N^{03}=N^{13}=N^{23}=0,  \notag \\
K_{L}^{0} &=&\beta K_{L}^{1}\Rightarrow N^{01}=x^{1}(\beta
K_{L}^{1}-K_{L}^{0})=0,  \label{osr} \\
K_{L}^{1} &=&qDx^{1},\ K_{L}^{2}=qDx^{2}\Rightarrow N^{02}=N^{12}=0.  \notag
\end{eqnarray}
Thus although in $S$ there is the 4D magnetic field (\ref{mg}) and there is
a part of $K_{L}$ (in $K_{L}^{2}$ in (\ref{si})), which corresponds to the
magnetic force, it is obtained that all components $N^{\mu \nu }$ are zero, $%
N^{\mu \nu }=0$, and consequently the whole torque $N=(1/2)N^{\mu \nu
}\gamma _{\mu }\wedge \gamma _{\nu }=0$. Every 4D CBGQ is invariant upon the
passive Lorentz transformations; the components transform by the Lorentz
transformations and the basis by the inverse Lorentz transformations leaving
the whole CBGQ unchanged. The invariance of some 4D CBGQ upon the passive
Lorentz transformations reflects the fact that such mathematical, invariant,
geometric 4D quantity represents \emph{the same physical object} for
relatively moving observers. Due to the invariance of any 4D CBGQ upon the
passive Lorentz transformations $N$ \emph{will be zero} \emph{in all other
relatively moving inertial frames, }thus in $S^{\prime }$, as well
\begin{equation}
N=(1/2)N^{\mu \nu }\gamma _{\mu }\wedge \gamma _{\nu }=(1/2)N^{\prime \mu
\nu }\gamma _{\mu }^{\prime }\wedge \gamma _{\nu }^{\prime }=0.  \label{enc}
\end{equation}
The paradox does not appear since the principle of relativity is
automatically satisfied in such an approach to special relativity which
exclusively deals with 4D geometric quantities, i.e., AQs or CBGQs, whereas
in the standard approach to special relativity$^{6}$ the principle of
relativity is postulated outside the framework of a mathematical formulation
of the theory.

The conclusion that can be drawn from this proof that $N$ is zero in all
relatively moving inertial frames is that the real cause of the violation of
the principle of relativity and of the existence of the paradox is the use
of 3D quantities as physical quantities in the 4D spacetime.

In the geometric approach to special relativity \emph{the independent
physical reality, both theoretically and experimentally, is attributed only
to the 4D geometric quantities, AQs or CBGQs, and not, as usual, to the 3D
quantities. }In Ref. 1 even the covariant quantities, e.g., $M^{\mu \nu }$, $%
x^{\mu }$, $u^{\nu }$, $F^{\alpha \beta }$, etc. are considered as auxiliary
mathematical quantities from which ``physical'' 3D quantities are deduced.

It is worth noting that the comparison$^{7}$ with well-known experiments
that test special relativity as are the Michelson-Morley experiment, the
''muon'' experiments, the Kennedy-Thorndike type experiments and the
Ives-Stilwell type experiments explicitly shows that all these experiments
are in a complete agreement with such an approach with 4D geometric
quantities, whereas, contrary to the general belief, it is not the case for
the usual approach that deals with, e.g., the Lorentz contraction and the
dilatation of time; the spatial distances and temporal distances taken
separately are not well-defined quantities in the 4D spacetime.\bigskip
\medskip

\noindent \textbf{REFERENCES\bigskip }

\noindent $^{1}$J.D. Jackson, ``Torque or no torque? Simple charged particle
motion observed

in different inertial frames,'' Am. J. Phys. \textbf{72,} 1484-1487 (2004).

\noindent $^{2}$T. Ivezi\'{c}, ``Torque or no torque?! The resolution of the
paradox using 4D

geometric quantities with the explanation of the Trouton-Noble experiment,''

physics/0505013; submitted to a research journal.

\noindent $^{3}$D. Hestenes, ``Spacetime physics with geometric algebra,''
Am. J Phys. \textbf{71},

691-714 (2003).

\noindent $^{4}$D.A. T. Vanzella, G.E.A. Matsas, H.W. Crater, ``Comment on
``General

covariance, the Lorentz force, and Maxwell equations,'' by H. W. Crater

[Am. J. Phys. \textbf{62} (10), 923-931 (1994)],'' Am. J. Phys. \textbf{64},
1075-76 (1996).

\noindent $^{5}$D. Hestenes and G. Sobczyk, \textit{Clifford Algebra to
Geometric Calculus}

(Reidel, Dordrecht, 1984).

\noindent $^{6}$A. Einstein, ``Zur Elektrodynamik bewegter K\"{o}rper,''
Ann. Physik. (Leipzig),

\textbf{17}, 891-921 (1905), tr. by W. Perrett and G.B. Jeffery, in \textit{%
The Principle of}

\textit{Relativity,} (Dover, New York, 1952).

\noindent $^{7}$T. Ivezi\'{c}, ``An invariant formulation\ of\ special\
relativity, or the ``True

transformations relativity,'' and comparison with experiments,'' Found.

Phys. Lett. \textbf{15} 27-69 (2002);

\end{document}